\documentclass[10pt, hidelinks]{article}
\usepackage[nocrop,newcenter,frame]{pagedims}
\usepackage{epsfig}
\usepackage{stata}
\usepackage{amsmath} 
\usepackage{authblk}
\usepackage{hyperref}
\usepackage{natbib}
\usepackage{supertabular}

\begin{document}

\title{PPMLHDFE: Fast Poisson Estimation with High-Dimensional Fixed Effects}


\author[1]{Sergio Correia}
\author[2]{Paulo Guimar\~aes}
\author[3]{Tom Zylkin}
\affil[1]{Federal Reserve Board, \href{mailto:sergio.a.correia@frb.gov}{sergio.a.correia@frb.gov}}
\affil[2]{Banco de Portugal, \href{mailto:pfguimaraes@bportugal.pt}{pfguimaraes@bportugal.pt}}
\affil[3]{University of Richmond,  \href{mailto:tzylkin@richmond.edu}{tzylkin@richmond.edu}}


\maketitle

\begin{abstract}
In this paper we present {\tt ppmlhdfe}, a new Stata command for estimation of (pseudo) Poisson regression models with multiple high-dimensional fixed effects (HDFE). Estimation is implemented using a modified version of the iteratively reweighted least-squares (IRLS) algorithm that allows for fast estimation in the presence of HDFE. Because the code is built around the {\tt reghdfe} package, it has similar syntax, supports many of the same functionalities, and benefits from {\tt reghdfe}'s fast convergence properties for computing high-dimensional least squares problems. Performance is further enhanced by some new techniques we introduce for accelerating HDFE-IRLS estimation specifically. {\tt ppmlhdfe} also implements a novel and more robust approach to check for the existence of (pseudo) maximum likelihood estimates.
\end{abstract}

Keywords: {\tt ppmlhdfe}, {\tt reghdfe}, Poisson regression, high-dimensional fixed-effects

\section{Introduction}

Poisson regression is now well established as the standard approach to model count data. However, it is also gaining popularity as a viable alternative for estimation of multiplicative models where the dependent variable is nonnegative. Commonly, these models are estimated by linear regression applied to {a log-transformed dependent variable}. But, as with ordinary least squares (OLS), the only assumption required for consistency of the Poisson regression estimator is the correct specification of the conditional mean of the dependent variable \citep{Gourieroux1984}. In this setting, Poisson regression becomes Poisson pseudo maximum likelihood (PPML) regression. Gourieroux et al.'s results greatly extend the realm of application of Poisson regression because there is no need to specify a distributional assumption for the dependent variable and, therefore, application is no longer restricted to count data. This means that PPML can be applied to any dependent variable with nonnegative values without the need to explicitly specify a distribution for the dependent variable. Moreover, unlike the log-linear model, PPML regression provides a natural way to deal with zero values on the dependent variable. Yet another advantage of PPML regression versus log-linear regression is that in the presence of heteroskedasticity, the parameters of log-linearized models estimated by OLS are inconsistent \citep{SantosSilva2006}. In this context, the use of robust standard errors to mitigate concerns about heteroskedaticity will lead to incorrect inference because OLS estimators are not consistent in the first place.

The potential of PPML regression was recognized early in the spatial sciences by \cite{Davies1987}, who recommended using pseudo--likelihood methods instead of the more popular Poisson regression for the modeling of spatial flows. However, it was not until \cite{SantosSilva2006} that PPML really took off, particularly in the international trade literature. In that paper, the authors made an excellent case for the PPML model and posited it as the ideal estimator for gravity equations. At around the same time, \cite{Blackburn2007} questioned the use of the traditional OLS approach for estimation of the Mincerian wage regression and proposed the use of pseudo-maximum likelihood estimators such as PPML regression. His basic point was essentially the same---labor economists routinely estimate wage regressions on micro datasets using log-linear regression, disregarding the fact that heteroskedasticity may undermine the validity of the results. 
{A similar critique has also taken hold in the health economics literature, where} the usage of log-linear regression to model health-care expenditures and utilization has been questioned  \citep[for example, ][]{manning2001}. Here, the more obvious reason {for the adoption of PPML} is the inadequacy of the log-transformation to deal with the large number of zeros typical in these areas. 

In sum, in the presence of nonnegative data with possibly many zeros, if one wants to make minimal assumptions about the {distribution of the data}, then PPML seems like the safest bet. This situation is very likely to occur across many areas of research, particularly when working with highly granular data (for example, when modeling firm R\&D expenditures, patent citation counts, daily product store sales, number of doctor visits, firm credit volumes, number of auction bidders, and number of commuters across regions).

Nevertheless, in applied work many researchers still resort to log-linear regressions in contexts where PPML would be better justified. One possible explanation is the ease with which researchers can  estimate linear regressions that control for multiple fixed effects. The increasing availability of larger panel-type datasets, coupled with advances in estimation techniques for linear regression models with high-dimensional fixed effects (HDFE) has allowed researchers to control for multiple sources of heterogeneity. Stata users are familiar with the user-written package {\tt reghdfe}, programmed by one of the authors, which has become Stata's standard tool for estimating linear models with multiple HDFE. 

In this paper we show that PPML with HDFE can be implemented with almost the same ease as linear regression with HDFE. To this end, we present {\tt ppmlhdfe}, a new Stata command for fast estimation of Poisson regression models with HDFE. The {\tt ppmlhdfe} command is to Poisson regression what {\tt reghdfe} represents for linear regression in the Stata world---a fast and reliable command with support for multiple fixed effects. Moreover, {\tt ppmlhdfe} takes great care to verify the existence of a maximum likelihood solution, adapting the innovations and suggested approaches described in \cite{Correia2019a}. {It also introduces some novel acceleration techniques relative to existing algorithms for HDFE nonlinear estimation that eliminate some unnecessary steps and lead to faster computation of the parameters of interest.}

\section{Stata Commands for Estimation of Models with HDFE}
The Stata community has been particularly active in developing and implementing methods to handle regression models that include more than one HDFE. The first such command, {\tt a2reg}, was coded by Amine Ouazad and was made available in 2008. The program was basically a port of the FORTRAN code written by Robert Creecy for estimation of a linear regression model with two HDFE. The approach is detailed in \cite{ACK2002} and involves solving the least-squares system of normal equations directly by application of the iterative conjugate gradient algorithm. The program provided the exact solution for the coefficients of the regression, but it lacked basic functionalities such as the calculation of the associated standard errors and data checks for multicollinearity. At around the same time, \citep{Cornelissen2008} introduced the command {\tt felsdvreg} that, like {\tt a2reg}, was meant for estimation of a linear regression with two HDFE. Cornelissen used a clever decomposition of the design matrix to simplify estimation.  His command was able to produce estimates of the standard errors, but his approach was only successful for particular data configurations and likely to break for larger datasets.
A couple of years later, \cite{Guimares2010} discussed an alternative algorithm for estimation of models with HDFE. Used in conjunction with the Frisch-Waugh-Lovell theorem (FWL), the algorithm could estimate these models using a minimum amount of memory and made easy the calculation of regular or one-way clustered standard errors.
Following the publication of \cite{Guimares2010}, Johannes Schmieder made available the {\tt gpreg} command while Guimar{\~a}es produced the {\tt reg2hdfe} command. Both commands used the general algorithm proposed in \cite{Guimares2010} along with the FWL transformation. While {\tt gpreg} was generally faster, {\tt reg2hdfe} was able to handle larger data sets.

Later, Sergio Correia developed what is currently the state-of-the-art estimation command for linear regression models with HDFE. This command, {\tt reghdfe}, offered several major improvements over existing commands. First, the convergence algorithm at the core of {\tt reg2hdfe} was improved and written in Mata, making it faster, and with better convergence properties \citep{Correia2016}. Second, it supported multiple HDFE and their interactions, allowing for the full usage of factorial variable notation to control for the fixed effects. Other relevant improvements consisted of support for instrumental-variables and different variance specifications, including multi-way clustering, support for weights, and the ability to use all post-estimation tools typical of official Stata commands such as {\tt predict} and {\tt margins}.\footnote{For a complete set of features of the command see \href{http://scorreia.com/software/reghdfe/}{http://scorreia.com/software/reghdfe/}.} By all accounts {\tt reghdfe} represents the current state-of-the-art command for estimation of linear regression models with HDFE, and the package has been very well accepted by the academic community.\footnote{As of December 2018, it had more than 7,000 hits at SSC, making it the 14th most downloaded Stata package. Google Scholar shows more than 200 citations.}  

The fact that {\tt reghdfe} offers a very fast and reliable way to estimate linear regression models with HDFE has opened up the way for estimation of other nonlinear regression models with HDFE. This is because many nonlinear models can be estimated by recursive application of linear regression. An obvious example is the nonlinear models that can be estimated by the nonlinear least-squares algorithm.\footnote{For a discussion of this estimation method see \cite{Davidson1993}.} Another example is the iteratively reweighted least squares (IRLS) algorithm that was developed for estimation of generalized linear models (GLMs). This was the approach implemented by Guimar{\~a}es in the user-written Stata command {\tt poi2hdfe} developed for estimation of PPML regression with two HDFEs. The {\tt poi2hdfe} command is basically a "wrapper" around {\tt reghdfe} that implements  estimation of a Poisson regression model with two HDFE.
Because structural gravity applications often require PPML models with three sets of fixed effects, Tom Zylkin also made available to the Stata community a specialized command called {\tt ppml\_panel\_sg}, which extended an earlier algorithm described in \cite{Figueiredo2015} and also built on the capabilities of {\tt reghdfe} \citep{Larch2018}.

The command {\tt ppmlhdfe} discussed in this paper implements PPML estimation with multiple HDFE, offering the full functionality of factorial variables to control for fixed effects. Thus, it can estimate the same models as {\tt poi2hdfe} and {\tt ppml\_panel\_sg} as well as more sophisticated models with multiple or interacted fixed effects, including models with heterogeneous slopes.\footnote{The command can be installed directly from the Statistical Software Components (SSC) archive. The development version, as well as complementary material, may be found at the dedicated Github repository: \href{https://github.com/sergiocorreia/ppmlhdfe}{https://github.com/sergiocorreia/ppmlhdfe}.}

To our knowledge, there are at present three other packages recently made available in R that also permit the estimation of Poisson regressions with multiple levels of fixed effects–{\tt alpaca} \citep{Stammann2018}, {\tt FENmlm} \citep{Berge2018}, and {\tt glmhdfe} \citep{glmhdfe}. Of these, {\tt alpaca} is the most similar to {\tt ppmlhdfe} in that it combines a within-transformation step with a Newton-Raphson estimation algorithm roughly equivalent to the IRLS method used here.
{\tt FENmlm} also involves a combination of these two steps, but uses a nonlinear Gauss-Seidel method to update the fixed effects \`a la  \cite{Figueiredo2015}.\footnote{To put the comparison another way, our method consists of within-transforming the data to take care of the fixed effects, using weighted least squares to update the remaining non-fixed effect coefficients (i.e., ``$\mathbf{\beta}$'' in what follows), updating the weights, and repeating. In \cite{Berge2018}, the within-transformation step is used to construct a concentrated Hessian for use in updating $\mathbf{\beta}$. A version of the concentrated Hessian appears in the weighted least squares step used here; thus, while the two algorithms are presented in different ways, they still share much in common.} 
{\tt glmhdfe}, meanwhile, does not use within-transformation and instead opts to embed Gauss-Seidel updating within IRLS, similar to the Stata command {\tt ppml\_panel\_sg.} One conceptual advantage our approach has over the former two methods is that we {devise a way to} solve the model without completely within-transforming the data {from scratch} in each iteration, thereby enabling us to realize significant speed gains. In addition, a general advantage of using a within-transformation approach over Gauss-Seidel is that it allows us to {easily handle models with heterogenous slopes as noted above}.

\section{Estimation Approach}

\subsection{The iteratively reweighted least squares algorithm}

GLMs are a class of regression models based on the exponential family of distributions that were introduced by \cite{Nelder1972}.
GLMs include popular nonlinear regression models such as logit, probit, cloglog, and Poisson. Following Ch. 12 of \cite{Hardin2018}, the exponential family is given by
\begin{equation}
f_y(y;\theta,\phi)= exp \left\{\frac {y\theta-b(\theta)}{a(\phi)}+c(y,\phi) \right\},
\end{equation}
where a(.), b(.), and c(.), are specific functions and $\phi$ and $\theta$ are parameters. For these models,
\begin{equation}
E(y)=\mu=b^{\prime}(\theta)
\end{equation}
and
\begin{equation}
V(y)=b^{\prime \prime}(\theta)a(\phi).
\end{equation}

Given a set of $n$ independent observations, each indexed by $i$, we can relate the expected value to a set of covariates ($\mathbf{x}_i$) by means of a link function g(.). More specifically it is assumed that
\begin{equation}
E(y_i)=\mu_i=g^{-1}(\mathbf{x}_i\mathbf{\beta}),
\end{equation}
and the likelihood for the GLM may be written as

\begin{equation}
L(\theta, \phi; y_1,y_2,...,y_n)= \prod\nolimits_{i=1}^{n} exp \left\{ \frac {y_i\theta_i-b(\theta_i)}{a(\phi)}+c(y_i,\phi) \right\}.
\end{equation}

Estimates for $\beta$ are obtained by solving the first-order conditions for maximization of the (pseudo) likelihood. Application of the Gauss-Newton algorithm with the expected Hessian leads to the following updating equation:

\begin{equation}
\mathbf{\beta}^{(r)}=\left (\mathbf{X}^{\prime}\mathbf{W}^{(r-1)}\mathbf{X} \right ) ^{-1}\mathbf{X}^{\prime}\mathbf{W}^{(r-1)}\mathbf{z}^{(r-1)} \label{IRLS},
\end{equation}
where $\mathbf{X}$ is the design matrix of explanatory variables, $\mathbf{W}^{(r-1)}$ is a weighting matrix,  
$\mathbf{z}^{(r-1)}$ is a transformation of the dependent variable, and $r$ is an index for iteration \citep[for details, see][]{Hardin2018}. Equation (\ref{IRLS}) makes it clear that the estimates of $\mathbf{\beta}$ are obtained by recursive application of weighted least squares. This approach is known as {Iteratively Reweighted Least Squares, or} IRLS.

\subsection{The Poisson regression model}

In the case of Poisson regression we have
\begin{equation}
E(y_i)=\mu_i=\exp(\mathbf{x}_i\mathbf{\beta)}
\end{equation}
and the regression weights to implement IRLS simplify to 
\begin{equation}
\mathbf{W}^{(r-1)}=\text{diag}
\left\{ \exp(\mathbf{x}_i\mathbf{\beta}^{(r-1)}) \right\} \label{HDFE-w}
\end{equation}
while the dependent variable for the intermediary regression becomes
\begin{equation}
z_i^{(r-1)}=\left\{ \frac{y-\exp(\mathbf{x}_i\mathbf{\beta}^{(r-1)})}{\exp(\mathbf{x}_i\mathbf{\beta}^{(r-1)})}+\mathbf{x}_i\mathbf{\beta}^{(r-1)}\right\}. \label{HDFE-z}
\end{equation}

Implementation of the IRLS updating regression in equation (\ref{IRLS}) requires only computation of the vector of fitted values  $\mathbf{x}_i\mathbf{\beta}^{(r-1)}$ obtained in the previous iteration. 

\subsubsection{Dealing with HDFE}

The difficulty of implementing IRLS in the presence of HDFE comes from the fact that $\mathbf{X}$ may contain a large number of fixed effects that render the direct calculation of $(\mathbf{X}^{\prime}\mathbf{W}^{(r-1)}\mathbf{X})$ impractical, if not impossible. The solution is to use an alternative updating formula that estimates only the coefficients of the non-fixed effect covariates (say, $\delta$), thus reducing the dimensionality of the problem. This is because equation (\ref{IRLS}) is a weighted linear regression and therefore we can rely on the FWL theorem to expurgate the fixed effects. This means that instead of equation (\ref{IRLS}), we can use the following updating equation:
\begin{equation}
\mathbf{\delta}^{(r)}=\left (\mathbf{\widetilde{X}}^{\prime}\mathbf{W}^{(r-1)}\mathbf{\widetilde{X}} \right )^{-1} 
\mathbf{\widetilde{X}}^{\prime}\mathbf{W}^{(r-1)}\widetilde{\mathbf{z}}^{(r-1)} \label{IRLS2},
\end{equation}
where $\mathbf{\widetilde{X}}$ and $\widetilde{\mathbf{z}}$ are weighted within-transformed versions of the main covariate matrix $\mathbf{X}$ and working dependent variable $\mathbf{z}$, respectively. Moreover, {the FWL theorem also implies that the residuals computed from equation \eqref{IRLS2} are the same as those from equation \eqref{IRLS}. This observation has two very useful implications for our purposes. First, it implies we can perform the needed updates to $\mathbf{W}$ and $\mathbf{z}$ using 
\begin{equation}
\mathbf{X} \mathbf{\beta}^{(r)} = \mathbf{z}^{(r-1)} - \mathbf{e}^{(r)},  \label{HDFE-IRLS}
\end{equation}
where  $\mathbf{e}^{(r)}$ is a vector collecting the residuals computed using \eqref{IRLS2}.
 New values for $\mathbf{W}^{(r)}$ and $\mathbf{z}^{(r)}$ then directly follow from \eqref{HDFE-w} and \eqref{HDFE-z}, as in the original IRLS loop.\footnote{%
A similar principle is also employed in \citet{Stammann2018}. Her formulation of the weighted least squares step differs in that she differences out the $\mathbf{x}_i \mathbf{\beta}^{(r-1)}$ term from the traditional IRLS dependent variable. This approach should nonetheless be roughly equivalent to IRLS computationally. Another difference between her algorithm and ours comes from the special acceleration techniques we have programmed into {\tt ppmlhdfe}, as we discuss next.} Second, it  also means that once $\delta^{(r)}$ converges to the correct estimate  $\mathbf{\hat\delta}$, the estimated variance-covariance matrix for the weighted least squares regression in \eqref{IRLS2} will be the correct variance-covariance matrix for $\mathbf{\hat\delta}$, and standard adjustments for heteroskedasticity and clustering similarly require no further special steps.}

\subsubsection{Accelerating HDFE-IRLS}

The user-written command {\tt poi2hdfe} implemented the updating equation ({\ref*{IRLS2}}) using {\tt reghdfe} as the workhorse for running the HDFE weighted least-squares regressions. This is a computationally intensive procedure requiring estimation of an HDFE regression model in every IRLS iteration. There are, however, several workarounds in {\tt ppmlhdfe} that make it much more efficient. For instance, {\tt ppmlhdfe} directly embeds the Mata routines of {\tt reghdfe}, thus taking advantage of the fact that some of the computations need to be done only once, as they remain the same for every IRLS iteration. But the most significant speed improvements come from the modifications {we have} introduced to the standard HDFE-IRLS algorithm aimed at reducing the number of calls to {\tt reghdfe}. {These modifications are as follows:}

{First, we within-transform (or ``partial out'') the original untransformed variables $\mathbf{z}$ and $\mathbf{X}$ in the first IRLS iteration only. From the second iteration onwards, we exploit the fact that, given an arbitrary linear combination of the fixed effects $\mathbf{d}$, partialing out $\mathbf{z} - \mathbf{d}$ is numerically equivalent to partialing out $\mathbf{z}$. Hence, if the eventual solution to the partial-out step is $\widetilde{\mathbf{z}}=\mathbf{z}^{(r)}-\mathbf{d}^{(r)}$, it is often much faster to partial out $\mathbf{z}^{(r)}-\mathbf{d}^{(r-1)}$  than it is to start from the untransformed $\mathbf{z}$ variable (since $\mathbf{d}^{(r-1)}$ is generally a reasonable initial guess for $\mathbf{d}^{(r)}$). In practice, we progressively
update $\widetilde{\mathbf{z}}$ by starting each within-transformation step using the new starting value  $\widetilde{\mathbf{z}}^{*(r)} = \widetilde{\mathbf{z}}^{(r-1)} + \mathbf{z}^{(r)} - \mathbf{z}^{(r-1)}$, where $\mathbf{z}^{(r)}$ and $\mathbf{z}^{(r-1)}$ are computed as in \eqref{IRLS}.\footnote{%
{Note that, if $\mathbf{d}^{(r)} \approx \mathbf{d}^{(r-1)}$, then $\widetilde{\mathbf{z}}^{(r)}=\mathbf{z}^{(r)}-\mathbf{d}^{(r)} \approx \mathbf{z}^{(r-1)}-\mathbf{d}^{(r-1)} + \mathbf{z}^{(r)} - \mathbf{z}^{(r-1)}=\widetilde{\mathbf{z}}^{(r-1)}+ \mathbf{z}^{(r)} - \mathbf{z}^{(r-1)}$. Moreover, since $\mathbf{X}$ is the same in every iteration it only needs to be partialled out in the initial iteration.}} We similarly are able to within-transform $\mathbf{{X}}$ in a progressive fashion by starting from the last values of $\mathbf{\widetilde{X}}$ in each iteration rather than starting over again from the original untransformed $\mathbf{X}$ variables.}

Second, another artifact that helps speed up convergence involves the choice of the criterion for the inner loops of {\tt reghdfe}. In our implementation, this criterion becomes tighter as we approach convergence, thus avoiding unnecessary {\tt reghdfe} iterations. {In sum, we are able to progressively within-transform both $\mathbf{X}$ and $\mathbf{z}$ while simultaneously updating the weights and $\mathbf{z}$ needed for the IRLS step, only requiring the within-transformation procedure to reach completion as the full algorithm converges. In practice, these innovations can reduce the total number of calls to {\tt reghdfe} by 50\% or more, leading to substantial speed gains in computation.}\footnote{For example, in the trade data example that follows, {\tt ppmlhdfe} at its default settings reaches convergence after {36} calls to {\tt reghdfe}. If we instead disable these features (by adding the options {\tt start\_inner\_tol(1e-08)} and {\tt use\_exact\_partial(1)}), the total number of calls increases to {98} (effectively an {170}\% increase in computing time). The exact same answer is achieved in either case.}

\subsubsection{Existence of MLE}

\cite{SantosSilva2010} and \cite{SantosSilva2011} noted that for some data configurations, maximum likelihood estimates for the Poisson regression may not exist. As a result, estimation algorithms may be unable to converge or may converge to the incorrect estimates. This situation bears some resemblance to the well-known problem of separation in the binary-choice model. In the case of Poisson regression, this happens if the log-likelihood increases monotonically as one or more coefficients tends to infinity. As shown by \cite{SantosSilva2010}, this may occur if there is multicollinearity among the regressors for the subsample of positive values of the dependent variable. To overcome this problem, they suggest identifying and dropping problematic regressors.\footnote{This approach is implemented in their user-written package {\tt ppml}.} However, which regressor(s) to drop is an ambiguous decision with implications {for} the identification of the remaining parameters. Moreover, in Poisson models with multiple HDFEs this strategy may not even be feasible.

In a recent paper \citep{Correia2019a}, we discuss the necessary and sufficient conditions for the existence of estimates in a wide class of GLM models and show that, in the case of Poisson regression, it is always possible to find MLE estimates if some observations are dropped from the sample. These observations---separated observations---do not convey relevant information for the estimation process and thus can be safely discarded. After dropping these observations, some regressors will become collinear and thus must also be dropped. Additionally, in the same paper, we propose a method to identify separated observations that will succeed even in the presence of HDFEs. By default, {\tt ppmlhdfe} implements this method (the \textbf{ir} method) plus three other methods to identify separated observations.\footnote{For a better understanding of the methods implemented in {\tt ppmlhdfe}, see our \href{https://github.com/sergiocorreia/ppmlhdfe/blob/master/guides/separation\_primer.md}{primer on separation} available on {\tt ppmlhdfe}'s Github repository.}

\section{The {\tt ppmlhdfe} Command}
\newcommand{\Ret}[2]{\stcmd{e(#1)} & #2}
\newcommand{\RetX}[2]{\stcmd{e(#1)} & \quad #2 \\}

{\tt ppmlhdfe} requires the installation of the latest versions of {\tt ftools} and {\tt reghdfe}.

\subsection{Syntax}

The syntax for {\tt ppmlhdfe} is similar to that of {\tt reghdfe}:

\begin{stsyntax}
	\dunderbar{}ppmlhdfe
	\depvar\
	\optindepvars\
	\optif\
	\optin\
	\optweight\
	, \optional{ \underbar{a}bsorb({\it absvars}) 
		\dunderbar{exp}osure(\varname)
		\underbar{off}set(\varname)
		d(\varname)
		d
		vce({\it vcetype\/})
		\underbar{v}erbose(\num)
		\underbar{nol}og(\num)
		\underbar{tol}erance(\num)
		guess(\ststring)
		\underbar{sep}aration(\ststring)
		\underbar{maxit}eration(\num)
		\underbar{keepsin}gletons
		{\it display\_options}
	}
\end{stsyntax}

\noindent
{\depvar} is the dependent variable. It must be nonnegative but it is not restricted to integer values. Use of time-series operators or factor variables (if they specify one level of the group) is allowed.

\noindent
\textit{indepvars} represents the set of explanatory variables in the regression. Both factor and time series operators are allowed.

\textbf{Options}

\noindent
{\tt \underbar{a}bsorb}(\textit{absvars} [, \textit{savefe}]) \textit{absvars} contains a list of all categorical variables that identify the fixed effects to be absorbed. Each variable represents one set of fixed effects. Factor variable notation can be used. If you want to save the estimates of the fixed effects, you can either assign a name to the new variable when specifying \textit{absvars}, as in \textit{newvar}=\textit{absvar}, or you can use the option \textit{savefe}, in which case all fixed effects estimates are saved using the stub \textit{\underline{{ }{ }}hdfe\num\underline{{ }{ }}}.

\hangpara
{\tt \underbar{exp}osure}(\varname) includes ln({\it varname}) in model with coefficient constrained to 1.

\hangpara
{\tt \underbar{off}set}(\varname) includes {\it varname} in model with coefficient constrained to 1.

\hangpara
{\tt d}(\varname) creates a new variable with the sum of the fixed effects. The option is required if you are absorbing fixed effects and planning on running {\tt predict} afterwards.

\hangpara
{\tt d} works as above but automatically names the variable \textit{\underline{{ }{ }}ppmlhdfe\underline{{ }{ }}}. 

\hangpara
{\tt vce(}{\it vcetype\/}{\tt )} where {\it vcetype} may be {\tt \underbar{r}obust} (default) or {\tt \underbar{cl}uster} {\it fvvarlist} (allowing two- and multi-way clustering).

\hangpara
{\tt \underbar{v}erbose}(\num) controls the amount of debugging information to show. Default is 0 but higher integer values will present increasing detail. The value -1 will prevent the displaying of any messages.

\hangpara
{\tt \underbar{nol}og}(\num) hides the iteration output.

\hangpara
{\tt \underbar{tol}erance}(\num) criterion for convergence. Default is 1e-8.

\hangpara
{\tt guess}(\ststring) rule for setting initial values; valid options are {\tt simple} (the default) and {\tt ols}.

\hangpara
{\tt \underbar{sep}aration}(\ststring) set rules for dropping separating observations; valid options are {\tt fe}, {\tt ir}, {\tt simplex}, and {\tt mu} (or any combination of those). By default the first three are used ({\tt fe simplex ir}). To disable all separation checks, set the option to {\tt none}. 

\hangpara
{\tt \underbar{maxit}eration}(\num) maximum number of iterations. The default is 10,000.

\hangpara
{\tt version} reports the current version and installed dependencies. Should not be used with any arguments. 

\hangpara
{\tt \underbar{keepsin}gletons} does not drop singleton groups.

\hangpara
{\tt \underbar{ef}orm} report exponentiated coefficients (incidence-rate ratios).

\hangpara
{\tt \underbar{ir}r} synonym for {\tt \underbar{ef}orm}.

More specialized options can be found in the \href{https://github.com/sergiocorreia/ppmlhdfe/blob/master/guides/undocumented.md}{documentation} available on the Github repository.  


\subsection{Saved results}
{\tt ppmlhdfe} saves the following results to \stcmd{e()}:

\begin{stresultsX}
	\stresultsgroup{Scalars} \\
	\RetX{N}{number of observations}
	\RetX{num\_singletons}{number of dropped singleton observations}
	\RetX{num\_separated}{number of separated observations}
	\RetX{N\_full}{number of observations, including dropped, singleton and separated obs.}
	\RetX{drop\_singletons}{1 if singleton obs. were searched for and dropped}
	\RetX{rank}{rank of \stcmd{e(V)}}
	\RetX{df}{residual degrees of freedom}
	\RetX{df\_m}{model degrees of freedom}
	\RetX{df\_a}{degrees of freedom lost due to the fixed effects}
	\RetX{df\_a\_initial}{number of categories in the fixed effects: \stcmd{e(df\_a)} - \stcmd{e(df\_a\_redundant)}}
	\RetX{df\_a\_redundant}{number of redundant fixed effect categories}
	\RetX{N\_hdfe}{number of absorbed fixed-effects}
	\RetX{N\_hdfe\_extended}{number of absorbed fixed-effects plus fixed-slopes}
	\RetX{rss}{residual sum of squares}
	\RetX{rmse}{root mean squared error}
	\RetX{chi2}{chi-squared}
	\RetX{r2\_p}{pseudo\_R\_squared }
	\RetX{ll}{log-likelihood}
	\RetX{ll\_0}{log-likelihood of fixed-effect-only regression}
	\RetX{N\_clustervars}{number of clustervars}
	\RetX{N\_clust\num}{number of clusters in the \num th cluster variable}
	\RetX{N\_clust}{number of clusters; minimum of all the \stcmd{e(clust\num)}}
	\RetX{ic}{number of iterations}
	\RetX{ic2}{number of iterations when partialling out fixed effects}
\\
\stresultsgroup{Macros} \\
	\RetX{cmd}{{\tt ppmlhdfe}}
	\RetX{cmdline}{command as typed}
	\RetX{separation}{list of methods used to detect and drop separated observations}
	\RetX{dofmethod}{dofmethod employed in the regression}
	\RetX{depvar}{name of dependent variable}
	\RetX{indepvars}{name of independent variables}
	\RetX{absvars}{name of the absorbed variables or interactions}
	\RetX{extended\_absvars}{expanded absorbed variables or interactions}
	\RetX{title}{title in estimation output}
	\RetX{clustvar}{name of cluster variable}
	\RetX{clustvar\num}{name of the {\num}th cluster variable}
	\RetX{vce}{{\it vcetype} specified in \stcmd{vce()}}
	\RetX{chi2type}{title used to label Std. Err.}
	\RetX{offset}{linear offset variable}
	\RetX{properties}{\textbf{b} \textbf{V}}
	\RetX{predict}{{\tt ppmlhdfe\_p}; program used to implement {\tt predict} }
	\RetX{estat\_cmd}{{\tt reghdfe\_estat}; program used to implement {\tt estat}}
	\RetX{marginsok}{predictions allowed by {\tt margins}}
	\RetX{marginsnotok}{predictions disallowed by {\tt margins}}
	\RetX{footnote}{{\tt reghdfe\_footnote}; program used to display the degrees-of-freedom table}
\\
\stresultsgroup{Matrices} \\
	\RetX{\textbf{b}}{coefficient vector}
	\RetX{\textbf{V}}{variance-covariance matrix of the estimators}
	\RetX{dof\_table}{number of categories, redundant categories, and degrees-of-freedom absorbed by each set of fixed effects}
\stresultsgroup{Functions} \\
	\RetX{sample}{marks estimation sample}
\end{stresultsX}

\section{Examples}

We start out with a very simple example that shows the advantage of {\tt ppmlhdfe}'s approach for dealing with the nonexistence of MLE estimates. As explained earlier, {\tt ppmlhdfe} takes great care to identify separated observations and then restricts the sample in a way that guarantees the existence of meaningful maximum-likelihood estimates. Our illustrative data consists of six observations and three explanatory variables:

\begin{stlog}
input y x1 x2 x3
0	1	2	1
0	0	0	2
0	2	3	3
1	1	2	4
2	2	4	5
3	1	2	6
end
\end{stlog}

If we try to estimate a Poisson regression with the {\tt glm} command, Stata fails to converge. The {\tt poisson} command produces estimates for all three coefficients associated with the Xs, but a quick inspection of results makes it clear that the results are unreliable:\footnote{A similar situation occurs if we estimate the Poisson regression using {\tt glm} with the \textit{irls} option.}

\begin{stlog}
	. poisson y x1 x2 x3, nolog
{\smallskip}
Poisson regression                              Number of obs     =          6
                                                LR chi2(3)        =       8.89
                                                Prob > chi2       =     0.0308
Log likelihood = -4.0415302                     Pseudo R2         =     0.5237
{\smallskip}
\HLI{13}{\TOPT}\HLI{64}
           y {\VBAR}      Coef.   Std. Err.      z    P>|z|     [95\% Conf. Interval]
\HLI{13}{\PLUS}\HLI{64}
          x1 {\VBAR}  -31.29521   8467.059    -0.00   0.997    -16626.43    16563.83
          x2 {\VBAR}   15.84339   4233.529     0.00   0.997    -8281.722    8313.408
          x3 {\VBAR}   .7970409   .4608654     1.73   0.084    -.1062388     1.70032
       _cons {\VBAR}  -4.032453   2.868066    -1.41   0.160    -9.653759    1.588853
\HLI{13}{\BOTT}\HLI{64}

\end{stlog}

The user-written command {\tt ppml} identifies the existence of a data problem and drops the variable \texttt{x3}. However, the fitted regression still shows problems:

\begin{stlog}
	. ppml y x1 x2 x3
{\smallskip}
note: checking the existence of the estimates
{\smallskip}
Number of regressors excluded to ensure that the estimates exist: 1
Excluded regressors:  x3
Number of observations excluded: 0
{\smallskip}
note: starting ppml estimation
{\smallskip}
Iteration 1:   deviance =  5.984675
(output ommitted)
Iteration 15:  deviance =  5.489052
Warning:  variance matrix is nonsymmetric or highly singular
{\smallskip}
Number of parameters: 3
Number of observations: 6
Pseudo log-likelihood: -6.5473014
R-squared: .3399843
Option strict is: off
WARNING: The model appears to overfit some observations with y=0
\HLI{13}{\TOPT}\HLI{64}
             {\VBAR}               Robust
           y {\VBAR}      Coef.   Std. Err.      z    P>|z|     [95\% Conf. Interval]
\HLI{13}{\PLUS}\HLI{64}
          x1 {\VBAR}  -32.31708          .        .       .            .           .
          x2 {\VBAR}   16.58591          .        .       .            .           .
       _cons {\VBAR}  -.8167293          .        .       .            .           .
\HLI{13}{\BOTT}\HLI{64}

\end{stlog}

If ran with the \textit{strict} option, {\tt ppml} will simply drop all regressors. The approach of {\tt ppmlhdfe} is quite different. Instead of searching for problematic regressors, it looks for problematic observations. In this case, {\tt ppmlhdfe} drops the third observation. It then drops \texttt{x2} to avoid perfect multicollinearity. The results are more plausible:\footnote{To our knowledge, besides {\tt ppmlhdfe}, no package in any other statistical software is capable of dealing with the separation problem in a robust way. Please see our \href{https://github.com/sergiocorreia/ppmlhdfe/blob/master/guides/separation\_benchmarks.md}{comparisons} available at the Github repository. In the site, we also provide a set of {\it csv} files which contain examples of separation that package developers can use to verify the robustness of their code.}

\begin{stlog}
	. ppmlhdfe y x1 x2 x3, nolog
(simplex method dropped 1 separated observation)
note: 1 variable omitted because of collinearity: x2
Converged in 6 iterations and 6 HDFE sub-iterations (tol = 1.0e-08)
{\smallskip}
PPML regression                                   No. of obs      =          5
                                                  Residual df     =          2
                                                  Wald chi2(2)    =      50.78
Deviance             =  .4775093816               Prob > chi2     =     0.0000
Log pseudolikelihood = -4.041530113               Pseudo R2       =     0.4532
\HLI{13}{\TOPT}\HLI{64}
             {\VBAR}               Robust
           y {\VBAR}      Coef.   Std. Err.      z    P>|z|     [95\% Conf. Interval]
\HLI{13}{\PLUS}\HLI{64}
          x1 {\VBAR}   .3914642   .1733026     2.26   0.024     .0517975     .731131
          x2 {\VBAR}          0  (omitted)
          x3 {\VBAR}   .7969293   .1582404     5.04   0.000     .4867838    1.107075
       _cons {\VBAR}  -4.031679   1.119578    -3.60   0.000    -6.226012   -1.837347
\HLI{13}{\BOTT}\HLI{64}

\end{stlog}
Next we replicate Example 1 shown on page 359 of the Stata 15 manual for the command \xtref{xtpoisson} with the fixed effects option. This example uses the {\it ships} dataset and estimates a Poisson regression of the number of {\it ship} accidents on several regressors. It treats the variable \textit{ship} as a fixed effect to control for five different types of ships. The regression is estimated with a control for exposure (\textit{service}) and the coefficients are reported as incidence rate-ratios. The syntax needed to replicate the example is \footnote{Note that the variable \textit{ship} is already set as the \textit{panelvar} in the ships dataset.}

\begin{stlog}
	. webuse ships, clear
{\smallskip}
. xtpoisson acc op_75_79 co_65_69 co_70_74 co_75_79, exp(service) irr fe nolog
{\smallskip}

\end{stlog}
To obtain equivalent results with {\tt ppmlhdfe}, we do
\begin{stlog}
	. ppmlhdfe acc op_75_79 co_65_69 co_70_74 co_75_79, a(ship) exp(service) irr nolog
Converged in 6 iterations and 6 HDFE sub-iterations (tol = 1.0e-08)
{\smallskip}
HDFE PPML regression                              No. of obs      =         34
Absorbing 1 HDFE group                            Residual df     =         25
                                                  Wald chi2(4)    =     111.06
Deviance             =  38.69505154               Prob > chi2     =     0.0000
Log pseudolikelihood = -68.28077143               Pseudo R2       =     0.8083
\HLI{13}{\TOPT}\HLI{64}
             {\VBAR}               Robust
    accident {\VBAR}     exp(b)   Std. Err.      z    P>|z|     [95\% Conf. Interval]
\HLI{13}{\PLUS}\HLI{64}
    op_75_79 {\VBAR}   1.468831   .1484359     3.80   0.000     1.204902    1.790572
    co_65_69 {\VBAR}   2.008002   .2202475     6.36   0.000     1.619572    2.489592
    co_70_74 {\VBAR}    2.26693   .3256501     5.70   0.000     1.710649    3.004107
    co_75_79 {\VBAR}   1.573695   .3117262     2.29   0.022     1.067358    2.320232
       _cons {\VBAR}   .0011254   .0001061   -72.03   0.000     .0009356    .0013538
 ln(service) {\VBAR}          1  (exposure)
\HLI{13}{\BOTT}\HLI{64}
{\smallskip}
Absorbed degrees of freedom:
\HLI{13}{\TOPT}\HLI{39}{\TRC}
 Absorbed FE {\VBAR} Categories  - Redundant  = Num. Coefs {\VBAR}
\HLI{13}{\PLUS}\HLI{39}{\RGTT}
        ship {\VBAR}         5           0           5     {\VBAR}
\HLI{13}{\BOTT}\HLI{39}{\BRC}

\end{stlog}
where we are absorbing \textit{ship} as a fixed effect. A few points are worth mentioning. As expected the estimated coefficients for the variables are the same as those obtained with the \xtref{xtpoisson} command. However, the results for the estimates of the standard errors are different because, by default, {\tt ppmlhdfe} reports robust standard errors.\footnote{If we used the option vce(robust) in the \xtref{xtpoisson} command, the results would still be different. This is because in \textbf{xtpoisson} (and other xt commands), Stata replaces vce(robust) with vce(cluster \textit{ships}), but does not apply a small-sample adjustment for the number of clusters (see [U]20.22).} 
The values for the log-likelihoods presented by the two commands are also different. The command \xtref{xtpoisson} reports the value of the conditional log-likelihood, while {\tt ppmlhdfe} reports the actual Poisson log-likelihood (and could thus possibly be used for likelihood ratio tests against a Poisson regression if one were willing to accept the Poisson distribution assumption). 
Given that we are working with a small dataset, we could replicate the results obtained with {\tt ppmlhdfe} using the \rref{poisson} command as in
\begin{stlog}
{\smallskip}
poisson acc op_75_79 co_65_69 co_70_74 co_75_79 i.ship, exp(service) irr vce(robust)
{\smallskip}
\end{stlog}
Note that we can absorb any categorical variable as a fixed effect. For example, if we were interested only in the coefficients for {\it op\_75\_79} and {\it co\_75\_79}, we could absorb {\it ship}, {\it co\_70\_74}, and {\it co\_75\_79} as

\begin{stlog}
. ppmlhdfe acc op_75_79 co_65_69, a(ship co_70_74 co_75_79) exp(service) irr nolog
Converged in 7 iterations and 23 HDFE sub-iterations (tol = 1.0e-08)
{\smallskip}
HDFE PPML regression                              No. of obs      =         34
Absorbing 3 HDFE groups                           Residual df     =         25
                                                  Wald chi2(2)    =      71.60
Deviance             =  38.69505154               Prob > chi2     =     0.0000
Log pseudolikelihood = -68.28077143               Pseudo R2       =     0.8083
\HLI{13}{\TOPT}\HLI{64}
             {\VBAR}               Robust
    accident {\VBAR}     exp(b)   Std. Err.      z    P>|z|     [95\% Conf. Interval]
\HLI{13}{\PLUS}\HLI{64}
    op_75_79 {\VBAR}   1.468831   .1484359     3.80   0.000     1.204902    1.790572
    co_65_69 {\VBAR}   2.008002   .2202475     6.36   0.000     1.619572    2.489592
       _cons {\VBAR}   .0015435   .0001204   -83.00   0.000     .0013247    .0017984
 ln(service) {\VBAR}          1  (exposure)
\HLI{13}{\BOTT}\HLI{64}
{\smallskip}
Absorbed degrees of freedom:
\HLI{13}{\TOPT}\HLI{39}{\TRC}
 Absorbed FE {\VBAR} Categories  - Redundant  = Num. Coefs {\VBAR}
\HLI{13}{\PLUS}\HLI{39}{\RGTT}
        ship {\VBAR}         5           0           5     {\VBAR}
    co_70_74 {\VBAR}         2           1           1     {\VBAR}
    co_75_79 {\VBAR}         2           1           1    ?{\VBAR}
\HLI{13}{\BOTT}\HLI{39}{\BRC}
? = number of redundant parameters may be higher

\end{stlog}
and the results would be exactly the same for the variables that were explicitly kept in the model.

Our third example provides a natural application of {\tt ppmlhdfe}. Here we estimate a gravity model using the ancillary data and example provided with the {\tt ppml\_panel\_sg} command. The dataset contains annual bilateral trade data for 35 countries from 1986 to 2004. The objective is to estimate the impact that \textit{fta}---a free trade agreement variable---has on trade. In this example we want to control for country pair fixed effects and country time fixed effects (for both importer and exporter countries). Additionally, we want our standard errors clustered at the level of the country-pair.
\begin{stlog}
	. use http://fmwww.bc.edu/RePEc/bocode/e/EXAMPLE_TRADE_FTA_DATA if category=="TOTAL", clear
(Example gravity data for ppml_panel_sg (35 countries, 1988-2004, every 4 yrs))
{\smallskip}
. cap egen imp=group(isoimp)
{\smallskip}
. cap egen exp=group(isoexp)
{\smallskip}
. ppmlhdfe trade fta, a(imp\#year exp\#year imp\#exp) cluster(imp\#exp) nolog
Converged in 11 iterations and 36 HDFE sub-iterations (tol = 1.0e-08)
{\smallskip}
HDFE PPML regression                              No. of obs      =      5,950
Absorbing 3 HDFE groups                           Residual df     =      1,189
Statistics robust to heteroskedasticity           Wald chi2(1)    =      21.04
Deviance             =  377332502.3               Prob > chi2     =     0.0000
Log pseudolikelihood = -188710931.7               Pseudo R2       =     0.9938
{\smallskip}
Number of clusters (imp\#exp)=      1,190
                            (Std. Err. adjusted for 1,190 clusters in imp\#exp)
\HLI{13}{\TOPT}\HLI{64}
             {\VBAR}               Robust
       trade {\VBAR}      Coef.   Std. Err.      z    P>|z|     [95\% Conf. Interval]
\HLI{13}{\PLUS}\HLI{64}
         fta {\VBAR}   .1924455   .0419527     4.59   0.000     .1102197    .2746713
       _cons {\VBAR}   16.45706   .0217308   757.32   0.000     16.41447    16.49965
\HLI{13}{\BOTT}\HLI{64}
{\smallskip}
Absorbed degrees of freedom:
\HLI{13}{\TOPT}\HLI{39}{\TRC}
 Absorbed FE {\VBAR} Categories  - Redundant  = Num. Coefs {\VBAR}
\HLI{13}{\PLUS}\HLI{39}{\RGTT}
    imp\#year {\VBAR}       175           0         175     {\VBAR}
    exp\#year {\VBAR}       175           5         170     {\VBAR}
     imp\#exp {\VBAR}      1190        1190           0    *{\VBAR}
\HLI{13}{\BOTT}\HLI{39}{\BRC}
* = FE nested within cluster; treated as redundant for DoF computation

\end{stlog}
Notice that we have used Stata factorial notation when specifying the three variables to be absorbed, which is generally faster than creating the categorical variables for the interactions beforehand.  

Because our trade example includes high-dimensional fixed effects, it also gives us an opportunity to unpack the key mechanisms behind how {\tt ppmlhdfe} works from a programming perspective. {\tt ppmlhdfe} itself is a complex program that is mainly coded in Mata and also takes advantage of the existing inner workings of {\tt reghdfe} whereever possible. But the essential algorithm used to implement HDFE-IRLS is quite portable and can be programmed in any language that already offers robust algorithms for within-transformation and standard weighted regression. The following simple Stata code helps to illustrate\footnote{This code is \href{https://github.com/sergiocorreia/ppmlhdfe/blob/master/examples/algorithm.do}{available} at the Github repository.}.

\begin{stlog}
use http://fmwww.bc.edu/RePEc/bocode/e/EXAMPLE_TRADE_FTA_DATA if category=="TOTAL", clear
egen imp = group(isoimp)
egen exp = group(isoexp)
egen pair = group(isoexp isoimp)
local accelerate = 1
local crit 1
local iter 0
local last .
local inner_tol = 1e-4
{\smallskip}
while (`crit' > 1e-8 | `inner_tol' > `crit') {\lbr}
    loc ++iter
    di as text _n "Iteration `iter' (crit=`crit') (HDFE tol=`inner_tol')"
    if (`iter'==1) {\lbr}
        qui su trade, mean
        qui gen double mu = (trade + r(mean)) / 2
        qui gen double eta = log(mu)
        qui gen double z = eta + trade / mu - 1
        qui gen double last_z = z
        qui gen double reg_z = z
    {\rbr}
    else if (`accelerate') {\lbr}
        qui replace last_z = z
        qui replace z = eta + trade / mu - 1
        qui replace reg_z = z - last_z + z_resid
        qui replace fta = fta_resid
    {\rbr}
    else {\lbr}
        qui replace z = eta + trade / mu - 1
        qui replace reg_z = z
    {\rbr}
    * Tighten HDFE tolerance
    if (`crit' < 10 * `inner_tol') {\lbr}
        local inner_tol = `inner_tol' / 10
    {\rbr}
    cap drop *resid
    * Perform HDFE Weighted Least Squares
    qui reghdfe reg_z [aw=mu], a(imp\#year exp\#year imp\#exp) ///
    res(z_resid) keepsing v(-1) tol(`inner_tol')
    qui reghdfe fta [aw=mu], a(imp\#year exp\#year imp\#exp) ///
    res(fta_resid) keepsing v(-1) tol(`inner_tol')
    reg z_resid fta_resid [aw=mu], noconstant cluster(pair) nohead
    predict double resid, resid
   
    * Update eta = Xb; mu = exp(eta)
    qui replace eta = z - resid 
    qui replace mu = exp(eta)
    local crit = abs(_b[fta_resid] - `last')
    local last = _b[fta_resid]
 {\rbr}

\end{stlog}
As the while loop at the center of this example code converges, the final point estimate and standard error for \textit{fta} will be the same as those computed by {\tt ppmlhdfe} above. Some key operations to point out  are the crucial IRLS updating step---{\tt replace eta = z - resid}---and the acceleration step that allows us to avoid having to within-transform $\mathbf{z}$ from scratch in each iteration---{\tt replace z = z\_resid + z - z\_last}. The {\tt reghdfe} command, when used without any righthand-side covariates, may be used to perform each within-transformation step. Also notice that we progressively tighten the {\tt reghdfe} tolerance when we approach convergence; as we have noted, full within-transformation is only required for the final set of estimates.\footnote{For simplicity, this example code uses convergence of the estimated coefficient for \textit{fta} as the stopping criterion. However, in the complete algorithm we require full convergence of the deviance, as in standard IRLS implementations.} 

Our final example showcases how to combine {\tt ppmlhdfe} with the {\tt esttab} package to construct publication-quality regression tables. To construct the labels for the fixed effects, we use the {\tt estfe} command included with {\tt reghdfe}.
The example consists of a database of more than a million observations, containing all SSCI citations collected between 1991 and 2009 for 100,404 articles published in 170 economics journals between 1991 and 2006.\footnote{For a detailed description of the dataset see \cite{Cardoso2010}. The dataset can be downloaded from the Github package repository.} The database has an ID for journal, article, the number of authors, the two-digit main JEL classification code\footnote{The JEL code is a detailed system of classification for articles in economics.} (124 codes), the publication year, the type of article (proceeding, journal article, review, or note) and the number of citations collected in each year. Nearly 58 percent of all observations are zeros. Our dependent variable is the number of citations, and we will pretend that our main interest is understanding whether the number of authors in an article does in fact increase the number of citations once we control for a variety of controls. We run the following specifications:

\begin{stlog}
. use citations\_example, clear
{\smallskip}
. estimates clear
{\smallskip}
. ppmlhdfe cit nbaut, a(issn type jel2 pubyear)
{\smallskip}
. eststo
{\smallskip}
. ppmlhdfe cit nbaut, a(issn#c.year type jel2 pubyear)
{\smallskip}
. eststo
{\smallskip}
. ppmlhdfe cit nbaut, a(issn#year type jel2 pubyear)
{\smallskip}
. eststo
{\smallskip}
. estfe *, labels(type "Article type FE" jel2 "JEL code FE" pubyear "Publication year FE" ///
issn "ISSN FE" issn#c.year "Year trend by ISSN" issn#year "ISSN-Year FE")
\end{stlog}

The results are summarized in the following table produced by the {\tt esttab} command. They show that the number of authors has a very robust impact on the citation count.
\begin{stlog}
	. esttab, indicate(`r(indicate_fe)', labels("Yes" "")) b(3) se(3) varwidth(25) label ///
>         stat(N ll, fmt(\%12.0fc \%13.1fc)) se starlevels(* 0.1 ** 0.05 *** 0.01) compress
{\smallskip}
\HLI{64}
                                (1)          (2)          (3)   
                                cit          cit          cit   
\HLI{64}
nbaut                         0.190***     0.190***     0.189***
                            (0.003)      (0.003)      (0.003)   
{\smallskip}
Constant                      0.054***     0.081***     0.110***
                            (0.006)      (0.006)      (0.006)   
{\smallskip}
Article type FE                 Yes          Yes          Yes   
{\smallskip}
JEL code FE                     Yes          Yes          Yes   
{\smallskip}
Publication year FE             Yes          Yes          Yes   
{\smallskip}
ISSN FE                         Yes                             
{\smallskip}
Year trend by ISSN                           Yes                
{\smallskip}
ISSN-Year FE                                              Yes   
\HLI{64}
N                         1,083,701    1,083,701    1,080,051   
ll                        -1,714,495.9    -1,685,149.1    -1,655,106.1   
\HLI{64}
Standard errors in parentheses
* p<0.1, ** p<0.05, *** p<0.01

\end{stlog}
What are the exact differences between the regressions? In all three regressions we introduced a fixed effect for the type of article, the JEL code, and the publication year. In the first specification we treat the journal as any other fixed effect. However, in the second specification we are assuming that there is a trend in the number of citations that is specific to each journal. Finally, the last specification introduces even more flexibility and permits the existence of a year-specific impact for each journal. While this is simply an illustrative example of the capabilities of {\tt ppmlhdfe}, if taken with due care the analysis of the estimates of the absorbed effects may also reveal interesting information.

\section{Conclusion}

{\tt ppmlhdfe} is a new user-written command that allows for fast estimation of (pseudo) Poisson regression models. It has a similar syntax to {\tt reghdfe} and many of the same functionalities. Moreover, {\tt ppmlhdfe} takes great care to check for existence of maximum likelihood results {and introduces some promising new concepts for accelerating nonlinear estimation with high-dimensional covariates}. Finally, we note that the estimation approach of {\tt ppmlhdfe} could easily be extended to any other model from the GLM family.

\bibliographystyle{sj}
\bibliography{./ppmlhdfe}

\providecommand{\noopsort}[1]{} \providecommand{\printfirst}[2]{#1}
  \providecommand{\singleletter}[1]{#1} \providecommand{\switchargs}[2]{#2#1}
\ifnum 21=1 \def\bibname{Reference}
\else \def\bibname{References} \fi
\begin{thebibliography}{21}
\expandafter\ifx\csname natexlab\endcsname\relax\def\natexlab#1{#1}\fi
\expandafter\ifx\csname url\endcsname\relax
  \def\url#1{\texttt{#1}}\fi
\expandafter\ifx\csname urlprefix\endcsname\relax\def\urlprefix{URL }\fi

\bibitem[{Abowd et~al.(2002)Abowd, Creecy, and Kramarz}]{ACK2002}
Abowd, J., R.~Creecy, and F.~Kramarz. 2002.
\newblock Computing {P}erson and {F}irm {E}ffects {U}sing {L}inked
  {L}ongitudinal {E}mployer--{E}mployee {D}ata.
\newblock Technical Report 2002-06, U.S. Census Bureau.
\urlprefix\url{http://lehd.dsd.census.gov/led/library/techpapers/tp-2002-06.pdf.}
\bibitem[{Berg{\'e}(2018)}]{Berge2018}
Berg{\'e}, L. 2018.
\newblock Efficient {E}stimation of {M}aximum {L}ikelihood {M}odels with
  {M}ultiple {F}ixed-{E}ffects: {T}he R {P}ackage FENmlm.
\newblock {CREA} {D}iscussion {P}aper {S}eries, Center for Research in Economic
  Analysis, University of Luxembourg.
\urlprefix\url{https://EconPapers.repec.org/RePEc:luc:wpaper:18-13.}
\bibitem[{Blackburn(2007)}]{Blackburn2007}
Blackburn, M.~L. 2007.
\newblock {Estimating {W}age {D}ifferentials without {L}ogarithms}.
\newblock \emph{Labour Economics} 14(1): 73--98.

\bibitem[{Cardoso et~al.(2010)Cardoso, Guimar{\~a}es, and
  Zimmermann}]{Cardoso2010}
Cardoso, A.~R., P.~Guimar{\~a}es, and K.~F. Zimmermann. 2010.
\newblock Trends in {E}conomic {R}esearch: An {I}nternational Perspective.
\newblock \emph{Kyklos} 63(4): 479--494.

\bibitem[{Cornelissen(2008)}]{Cornelissen2008}
Cornelissen, T. 2008.
\newblock The {S}tata {C}ommand felsdvreg to {F}it a {L}inear {M}odel with
  {T}wo {H}igh-{D}imensional {F}ixed {E}ffects.
\newblock \emph{Stata Journal} 8(2): 170--189.

\bibitem[{Correia(2016)}]{Correia2016}
Correia, S. 2016.
\newblock reghdfe: {E}stimating {L}inear {M}odels with {M}ulti-{W}ay {F}ixed
  {E}ffects.
\newblock 2016 {S}tata conference, Stata Users Group.

\bibitem[{Correia et~al.(2019)Correia, Guimar{\~a}es, and
  Zylkin}]{Correia2019a}
Correia, S., P.~Guimar{\~a}es, and T.~Zylkin. 2019.
\newblock {Verifying the {E}xistence of {M}aximum {L}ikelihood {E}stimates for
  Generalized {L}inear {M}odels}.
\newblock Technical report, University of Richmond.

\bibitem[{Davidson and MacKinnon(1993)}]{Davidson1993}
Davidson, R., and J.~MacKinnon. 1993.
\newblock \emph{Estimation and Inference in Econometrics}.
\newblock NY: Oxford University Press.

\bibitem[{Davies and Guy(1987)}]{Davies1987}
Davies, R.~B., and C.~M. Guy. 1987.
\newblock The {S}tatistical {M}odeling of {F}low {D}ata when the {P}oisson
  {A}ssumption is {V}iolated.
\newblock \emph{Geographical Analysis} 19(4): 300--314.

\bibitem[{Figueiredo et~al.(2015)Figueiredo, Guimar{\~a}es, and
  Woodward}]{Figueiredo2015}
Figueiredo, O., P.~Guimar{\~a}es, and D.~Woodward. 2015.
\newblock {Industry {L}ocalization, {D}istance {D}ecay, and {K}nowledge
  {S}pillovers: {F}ollowing the {P}atent {P}aper {T}rail}.
\newblock \emph{Journal of Urban Economics} 89(C): 21--31.

\bibitem[{Gourieroux et~al.(1984)Gourieroux, Monfort, and
  Trognon}]{Gourieroux1984}
Gourieroux, C., A.~Monfort, and A.~Trognon. 1984.
\newblock {Pseudo Maximum Likelihood Methods: Theory}.
\newblock \emph{Econometrica} 52(3): 681--700.

\bibitem[{Guimar{\~a}es and Portugal(2010)}]{Guimares2010}
Guimar{\~a}es, P., and P.~Portugal. 2010.
\newblock {A {S}imple {F}easible {P}rocedure to {F}it {M}odels with
  {H}igh-{D}imensional {F}ixed {E}ffects}.
\newblock \emph{Stata Journal} 10(4): 628--649.

\bibitem[{Hardin and Hilbe(2018)}]{Hardin2018}
Hardin, J.~W., and J.~Hilbe. 2018.
\newblock \emph{Generalized Linear Models and Extensions, Fourth Edition}.
\newblock College Station, Texas: Stata Press.

\bibitem[{Hinz et~al.(2019)Hinz, Hudlet, and Wanner}]{glmhdfe}
Hinz, J., A.~Hudlet, and J.~Wanner. 2019.
\newblock Separating the Wheat from the Chaff: Fast Estimation of GLMs with
  High-Dimensional Fixed Effects.
\newblock Retrieved from https:/github.com/julianhinz/R\_glmhdfe.

\bibitem[{Larch et~al.(2019)Larch, Wanner, Yotov, and Zylkin}]{Larch2018}
Larch, M., J.~Wanner, Y.~V. Yotov, and T.~Zylkin. 2019.
\newblock Currency Unions and Trade: A PPML Re-assessment with High-dimensional
  Fixed Effects.
\newblock \emph{Oxford Bulletin of Economics and Statistics (forthcoming)} .
\urlprefix\url{https://onlinelibrary.wiley.com/doi/abs/10.1111/obes.12283.}
\bibitem[{Manning and Mullahy(2001)}]{manning2001}
Manning, W.~G., and J.~Mullahy. 2001.
\newblock Estimating {L}og {M}odels: to {T}ransform or not to {T}ransform?
\newblock \emph{Journal of Health Economics} 20(4): 461--494.

\bibitem[{Nelder and Wedderburn(1972)}]{Nelder1972}
Nelder, J., and R.~Wedderburn. 1972.
\newblock Generalized Linear Models.
\newblock \emph{Journal of the Royal Statistical Society, Series A} 135:
  370--384.

\bibitem[{{Santos Silva} and Tenreyro(2006)}]{SantosSilva2006}
{Santos Silva}, J. M.~C., and S.~Tenreyro. 2006.
\newblock {The Log of Gravity}.
\newblock \emph{The Review of Economics and Statistics} 88(4): 641--658.

\bibitem[{{Santos Silva} and Tenreyro(2010)}]{SantosSilva2010}
\mbox{\vrule width30.25006ptheight2.62222ptdepth-2.25222pt}. 2010.
\newblock {On the Existence of the Maximum Likelihood Estimates in Poisson
  Regression}.
\newblock \emph{Economics Letters} 107(2): 310--312.

\bibitem[{{Santos Silva} and Tenreyro(2011)}]{SantosSilva2011}
\mbox{\vrule width30.25006ptheight2.62222ptdepth-2.25222pt}. 2011.
\newblock {Poisson: Some Convergence Issues}.
\newblock \emph{Stata Journal} 11(2): 215--225.

\bibitem[{Stammann(2018)}]{Stammann2018}
Stammann, A. 2018.
\newblock Fast and Feasible Estimation of Generalized Linear Models with
  High-Dimensional k-way Fixed Effects.
\newblock \emph{ArXiv e-prints} .
\urlprefix\url{https://arxiv.org/pdf/1707.01815v2.pdf.}
\end{thebibliography}

\clearpage
\end{document}